\documentclass[12pt, letterpaper]{article}

\usepackage{mathptmx}
\usepackage{amsmath}
\usepackage{array}
\usepackage[pdftex]{graphicx}

\usepackage[caption=false,font=footnotesize]{subfig}
\usepackage{stfloats}

\usepackage{xcolor}
\usepackage[colorlinks=true,linkcolor={black},citecolor={blue}]{hyperref}

\usepackage{caption} 
\title{Competence-Based Student Modelling with Dynamic Bayesian Networks}
\author{Rafael~Morales-Gamboa
\thanks{Rafael~Morales-Gamboa is with the Virtual University System, University of Guadalajara, Guadalajara, Jalisco, 30332 Mexico.
E-mail: rmorales@suv.udg.mx} \and L.~Enrique~Sucar%
\thanks{L. Enrique Sucar is with the Department of Computing, National Institute for Astrophysics, Optics and Electronics, Tonantzintla, Puebla, 72840 Mexico.
E-mail: esucar@inaoep.mx}}

\begin{document}
\maketitle
\begin{abstract}
We present a general method for using a competences map, created by defining generalization/specialization and inclusion/part-of relationships between competences, in order to build an overlay student model in the form of a dynamic Bayesian network in which conditional probability distributions are defined per relationship type. We have created a competences map for a subset of the transversal competences defined as educational goals for the Mexican high school system, then we have built a dynamic Bayesian student model as said before, and we have use it to trace the development of the corresponding competences by some hypothetical students exhibiting representative performances along an online course (low to medium performance, medium to high performance but with low final score, and two terms medium to high performance). The results obtained suggest that the proposed way for constructing dynamic Bayesian student models on the basis of competences maps could be useful to monitor competence development by real students in online course.
	
\end{abstract}
	


\section{Introduction}
\label{sec:intro}

%
%
%
%
Competences
have grown in popularity in the western educational world \cite{gordonKeyCompetencesEurope2009,lurieDeconstructingCompetencybasedEducation2017,nunezcortesModeloCompetencialCompetencia2016}, and so the interest on developing computational models for competences that can be used to support a variety of educational processes, from creating digital catalogues of competences to course design to monitoring competence development by students. Although meaning varies among organisations, in this paper we will assume a definition of \emph{competence} along the line of `the capability of someone to act effectively in some kind of situations, which demands the mobilization
of a variety of internal and external resources' which broadly integrates aspects of external performance and internal composition of competences that emerge in the literature.

Research in this area is important because little information is available regarding what competences the students have developed along their studies, and to what extend, beyond the stated learning objectives of the educational programmes they are subscribed in, and the titles of the courses they have taken and passed. Furthermore, information regarding the development of competences do not accumulate, neither at school nor later in life. For example, transversal competences are develop along many courses on specific contexts (e.g. problem solving in mathematics, geography, or biology), as well as through experiences at work and social interactions, yet there is no much accumulation of evidence regarding their development, particularly in a way suitable to provide automated support for teaching, learning, certifying, applying for a job, or hiring someone.

As e-learning provides facilities for automatic recollection of information  regarding the development of competences by students, which are not equally available in face to face education, we have proposed to afford the digital e-learning environment with detailed information regarding competences, their interrelationships, and their relations to course activities, so that evidence of competence development by students can be accumulated, transformed into knowledge, and used to support the educational processes, particularly those related to decision making regarding the development of competences by students \cite{moralesIntelligentEnvironmentDistance2009}. More recently, we have proposed  generic mechanisms for creating probabilistic graphical models (Bayesian networks) to trace the development of competences by students on the basis of competences maps \cite{morales-gamboaProbabilisticRelationalLearner2017}.

In this paper we present initial results from using such mechanisms for building a student model as a dynamic Bayesian network, and using it to trace the development of corresponding competences by some hypothetical students exhibiting prototypical performances. The estimates calculated by the network were compared against estimates provided by a sample of teachers. The results strongly suggest correlations among both sets of estimates, yet the teachers seemed to be more optimistic, and certain, about the development of competences by students, as if they were assuming, in the absence of evidence, competence improvement rather than decay along time. Furthermore, teachers seemed to value the previous history of evidence much less than the actual one.

We proceed by firstly providing a summary of related work in \autoref{sec:related}, followed by an explanation of how we construct our competences maps, in \autoref{sec:maps}. Then, in \autoref{sec:dbn} we describe with some detail how a dynamic Bayesian network is generated from the competences maps, including how the conditional probabilistic distributions are constructed for each type of relationship and, in some cases, for some specific ones. \autoref{sec:models} is devoted to describe the method followed for generating estimates of competence levels developed by prototypical students, both using the dynamic Bayesian network presented in previous sections, and through a questionnaire responded by a sample of teachers. The results obtained are compared in \autoref{sec:comparison}, and we provide some conclusions and suggestions for future work in \autoref{sec:conclusions}.

\section{Related work}
\label{sec:related}

\noindent There has been a considerable amount of work on computational representations for competences in this century. There is a standard \cite{competency_data_working_group_1484.20.1-2007_2008} and a recommendation \cite{imsgloballearningconsortiumIMSReusableDefinition2002} on how to encode competences for exchange between applications, which emphasise detailed description of competences in terms of their components, but they also include basic facilities for establishing relationships between competences. There has been work on extending the recommendation in order to have better descriptions of relationships between competetences, as well as ways for encoding and exchanging levels of competences development \cite{sampsonDevelopingCommonMetadata2007}. Further work exists on providing detailed descriptions of concepts, skills, procedures, principles and other competence elements, as well as including composition and specialization relationships, and complex procedures and conceptual structures, what enables complex descriptions of competences \cite{paquetteLearningDesignBased2006} and computational tools to deal with them \cite{stoofWebbasedSupportConstructing2007}. There is also work on developing rather formal definitions of competences for knowledge management inside an organization \cite{pepiotUECMLUnifiedEnterprise2007}. More recently, there is again a proposal of extending the recommendation commented previously \cite{imsgloballearningconsortiumIMSReusableDefinition2002}, this time to include relationships among competences as part of the model, distinguishing between general and specific competences, in the sense of performances in domains and subdomains, hence proposing a general framework for describing competences maps \cite{elasameCompetencyModelReview2018}.

Regarding the use of (dynamic) Bayesian networks for student modelling, a 2010 study \cite{conatiBayesianStudentModeling2010} suggest it has been motivated (1) by the large amount of uncertainty in estimating the cognitive or affective state of students on the basis of observations of their behaviour, and other measurable information, (2) by their sound foundations on probability theory to carry out inferences, and (3) by their transparency in comparison with other numerical representations such as neural networks. Nevertheless, a key issue in (dynamic) Bayesian student modelling is the simplification of the domain by establishing conditional independence between nodes, as well as the definition of the conditional probability distributions, as numerical representations of the influences of the parents of conditionally dependent nodes in the network, a task that can become daunting as the network grows and usually demands the availability of a large amount of data and the use of machine learning techniques \cite{kaserDynamicBayesianNetworks2017,sucarProbabilisticGraphicalModels2015a}. Recent work \cite{kaserDynamicBayesianNetworks2017} demonstrates a generic way of constructing student models as dynamic Bayesian networks on the basis of ``skill topologies'', defined by prerequisite relationships between skills, and a large collection of data.

So, the main contributions of this work are, on one hand, the integration of two fields of research that have been explored somehow separately: computational representations of competences maps and (dynamic) Bayesian student modelling; on the other hand, the proposal of a method for defining conditional probability distributions on the basis of types of relationships in competences maps of the kind shown in \autoref{sec:maps}, which go beyond the prerequisite type. 

\begin{figure*}[!t]
	\centering
	\includegraphics[width=5.0in]{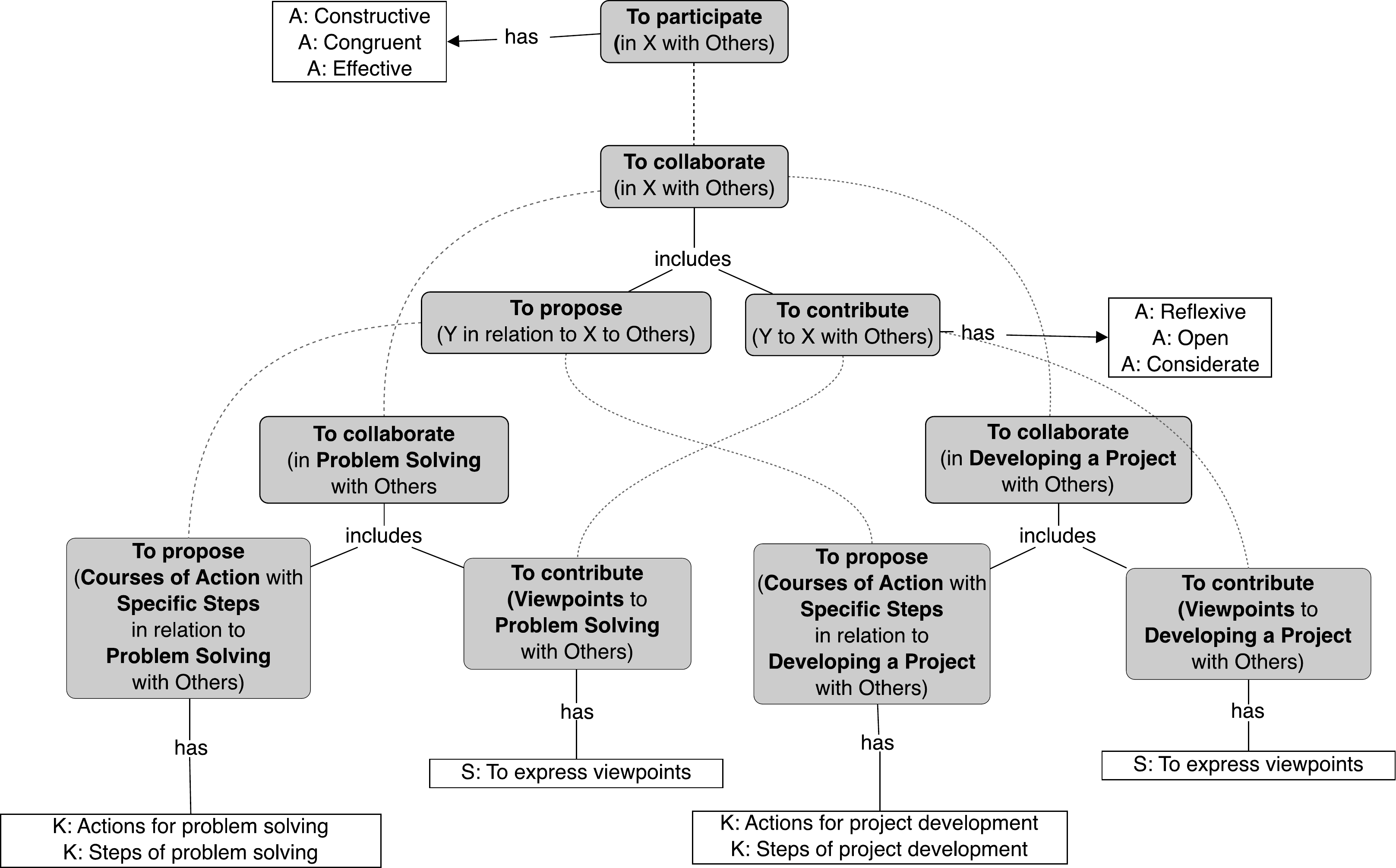}
	\caption{Competences map for `To participate and to collaborate effectively in diverse teams'. Competences are shown in rounded yellow boxes, whereas attributes are shown in plain white boxes. Generalizacion/specialization relationships are shown as dashed lines, whereas inclusion/part-of relationships are shown as solid lines labelled with the `includes' tag. The relationship between competences and attributes are labelled with the `has' tag.}
	\label{fig:competenceMap}
\end{figure*}

\section{Competence maps}
\label{sec:maps}

Our work is based on the notion of \emph{competence} as the capability to carry out a given action in a given context through the mobilization of various cognitive, affective, and conative resources, such as knowledge, skills, attitudes and values \cite{chanCompetencyAnalyserKnowledgebased2010}. This definition allow us to define two kinds of relationships between competences: \emph{generalization/specialization} relationships, generated through the removal or addition of resources (we call them \emph{competence attributes}), respectively, and the \emph{inclusion/part-of} relationship, generated by considering the attributes of some competences being part of a larger one, or by distributing the resources of a given competence along some sub-competences. We call a collection of competences interrelated by relationships of these kinds a \textit{competences map}.

In order to evaluate the expressiveness of our formalism, we have applied it to model the set of transversal competences established as a central element of the learning objectives of the National High School System in Mexico \cite{medinafloresMapaCompetenciasGenericas2017,secretariadeeducacionpublicaACUERDONumero4442008}. For example, the eighth competence in the set of transversal competences is described in the official documentation \cite{secretariadeeducacionpublicaACUERDONumero4442008} as follows:

\begin{quotation}
	\noindent To participate and to collaborate effectively in diverse teams.
	
	\noindent \textit{Attributes}:
	\begin{itemize}
		\item To propose ways to solve a problem or develop a team project, defining
		a course of action with specific steps.
		\item To provide points of view with openness and to consider those of other people in a reflective manner.
		\item To assume a constructive attitude, congruent with their knowledge and skills, within different work teams.
	\end{itemize}
\end{quotation}

By applying the formalism briefly described above (more details can be found in \cite{morales-gamboaProbabilisticRelationalLearner2017}) we generate the competences map presented in \autoref{fig:competenceMap} which includes the definition of the large and generic competence `to collaborate' as decomposed into two sub-competences, equally generic, `to propose' and 'to contribute'; a generic structure that is then specialized into collaborating in problem solving and collaborating in project execution. The full set of transversal competences for high school includes eleven competences, but the application of our formalism identifies much more, as illustrated in \autoref{fig:competenceMap}; over a hundred competences (nodes), considering both generic competences and more specific ones.

\begin{figure}[!t]
	\centering
	\includegraphics[width=2.5in]{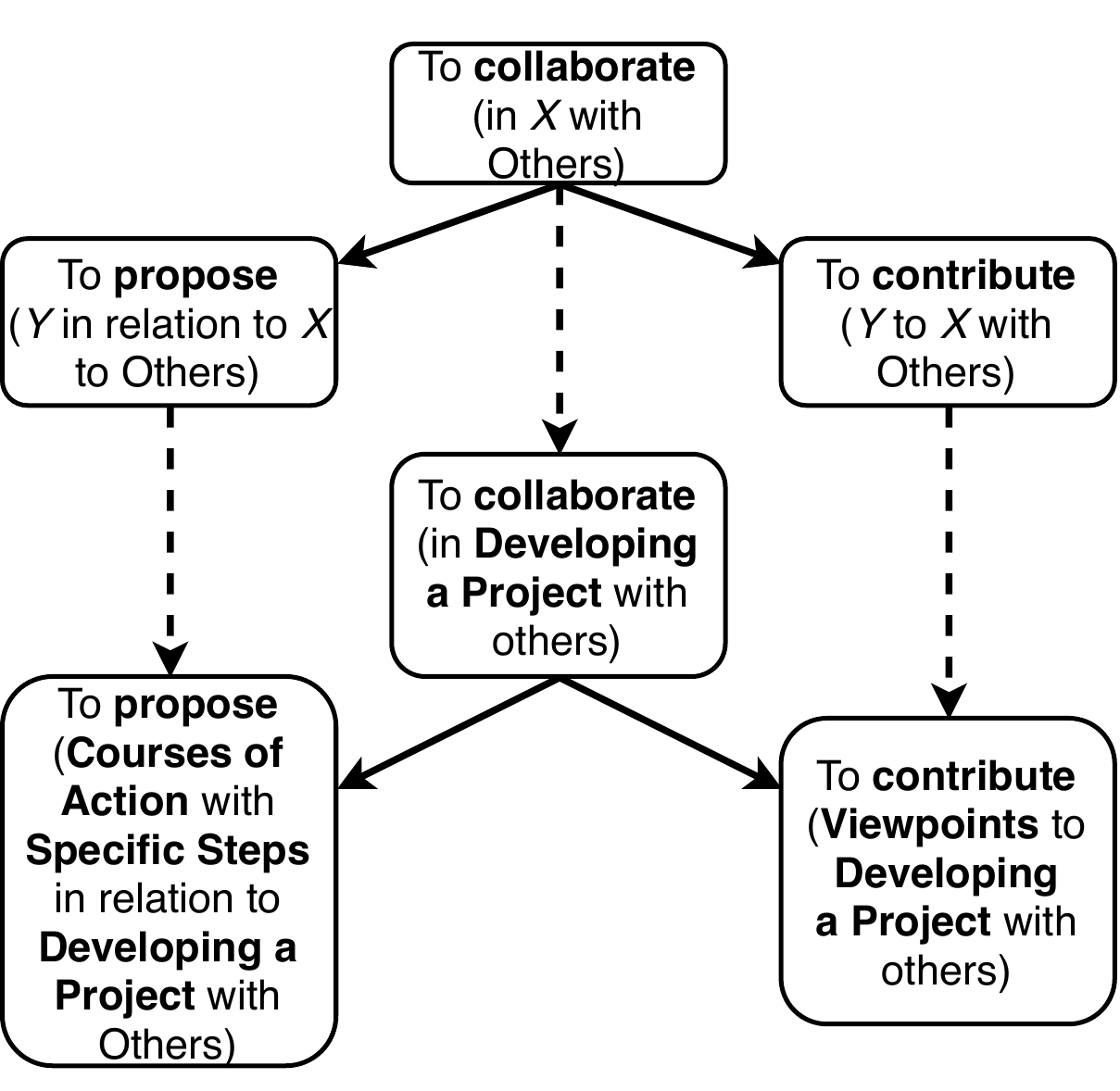}
	\caption{Submap of the competences map shown in \autoref{fig:competenceMap} that is used in the study presented in this paper. Generalizacion/specialization relationships are shown as dashed lines, whereas inclusion/part-of relationships are shown as solid lines.}
	\label{fig:projectCollaborationMap}
\end{figure}

In order to simplify the competences map for the purposes of this study, by eliminating the repetitions included in the map shown in \autoref{fig:competenceMap}, in the rest of the paper we will use the submap shown in \autoref{fig:projectCollaborationMap} that focus on collaborative project development.

\section{Dynamic Bayesian networks}
\label{sec:dbn}

\noindent A competence can be observed only through performances in concrete situations, and such performances are considered evidence of the level of competence. In the same way, the development of a more specific competence is evidence of the development of a more general one—as they share some key attributes. In a similar way, development of a super-competence cannot occur independently of the development of its sub-competences. So we could attribute some kind of causality to the generalization/specialization and inclusion/part-of kinds of relationships, at least in the case of competences. We then propose to create overlay student models \cite{holtStateStudentModelling1994,vanlehnStudentModelling1988} to trace the development of competences by students, associating a belief on the degree of development to each competence in the map, and transforming the competences map into a Bayesian network \cite{sucarProbabilisticGraphicalModels2015a}. As student competences evolve along time on the basis of their previous levels and new learning experiences, beliefs about their new levels are dependent both on beliefs about their previous levels and new evidence, so the Bayesian network should be dynamic. Furthermore, we assume that a new level of any competence is independent from old levels of the other competences in the map given the current level of such competence, and so are beliefs.

\begin{figure*}[!t]
	\centering
	\includegraphics[width=5.0in]{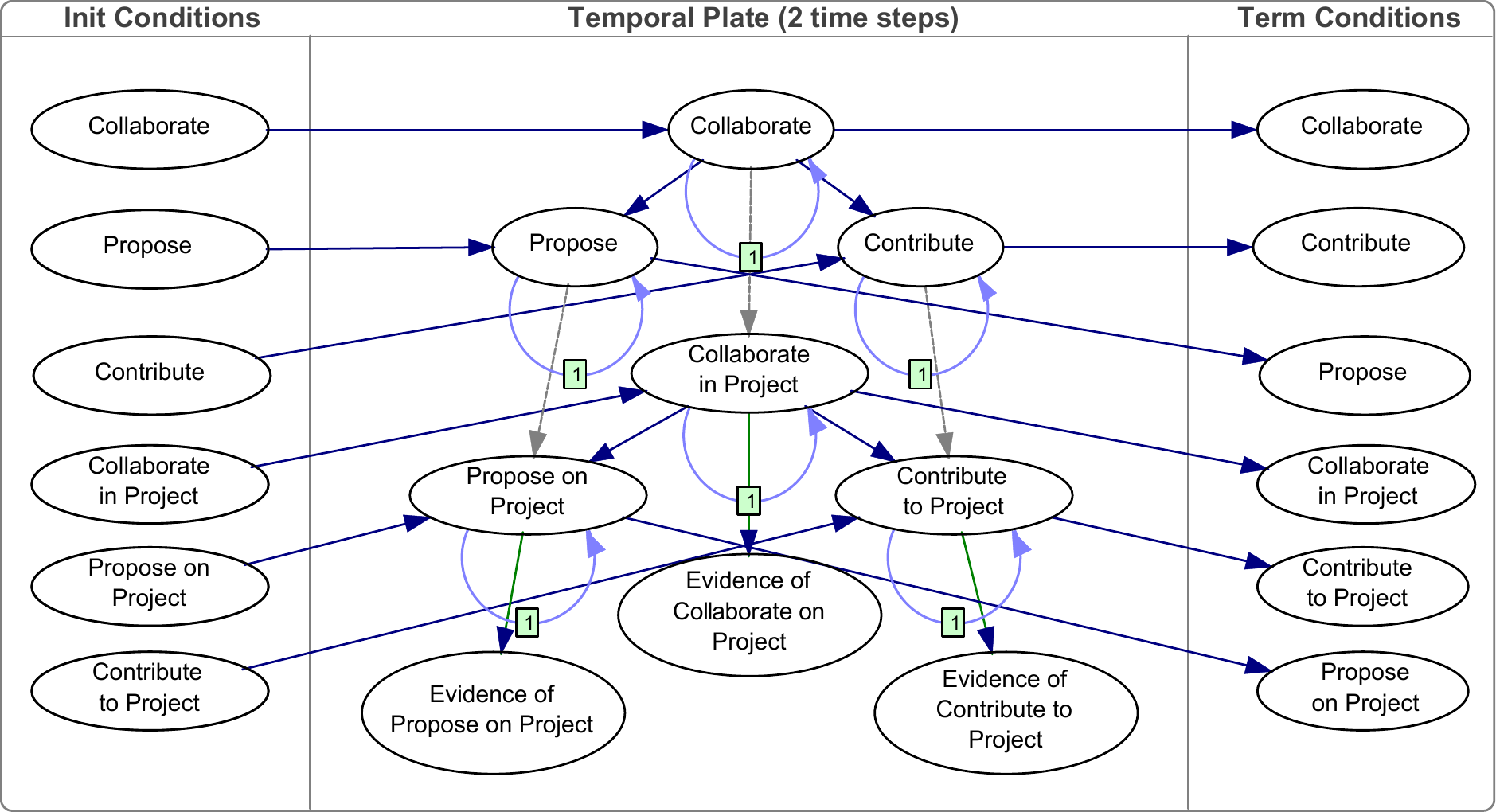}
	\caption{Dynamic Bayesian network corresponding to the competences map in \autoref{fig:projectCollaborationMap}. The nodes on the left side (Init Conditions) are initialized with the previous state of beliefs. The nodes in the central area (Temporal Plate) are firstly influenced by the nodes on the left, so they reproduce the previous state of the network; then they move one step in time, taking the previous state of the network and new evidence into account. Finally, the new beliefs are transferred to the nodes on the right (Term Conditions), from where they can be recovered.}
	\label{fig:collaborationDBN}
\end{figure*}

The dynamic Bayesian network (DBN) corresponding to the competences map presented in \autoref{fig:projectCollaborationMap} is shown in \autoref{fig:collaborationDBN}. \emph{Init(ial) Conditions} nodes are set to the beliefs on the previous levels of the competences—e.g. recovered from a database— while nodes in \emph{Term(inal) Conditions} are used to recover the new beliefs, on the current level of the competences. The nodes in the \emph{Temporal Plate} actually stand for two instances of the (non dynamic) Bayesian network built from the competences map, which stand for two time steps and are linked with temporal relationships (shown as round arrows in the figure) between equivalent nodes; in addition, the second instance includes nodes for evidences, at the bottom. The first instance is linked to the \emph{Init(ial) Conditions}, whereas the second instance is linked to the \emph{Term(inal) Conditions}.

The operation of the proposed dynamic Bayesian student modelling based on competences maps is then as follows:
\begin{enumerate}
	\item Given a student, and \emph{any} competences map composed as described in \autoref{sec:maps}, a non dynamic Bayesian network is built from the competences map using the conditional probability distributions described in \autoref{sec:conditionals}. Then flat probability distributions are set on the top nodes, and propagated. The final states of the beliefs in the nodes of the network are then stored somewhere as the initial, and current, level of the dynamic Bayesian network ($t=0$).
	\item When new evidence arrives at time $t=k$ ($k$ an integer greater than zero), the following process is iterated $k$ times:
	\begin{enumerate}
		\item The nodes in the \emph{Init(ial) Conditions} are set to the current beliefs, which correspond to the previous levels of the competences.
		\item The beliefs in the \emph{Init(ial) Conditions} are propagated to the first instance of the (non dynamic) Bayesian network.
		\item The beliefs in the first instance of the (non dynamic) Bayesian network are propagated to the   second instance using the temporal relationships, together with the evidence in the bottom nodes (if $t = k$).
		\item The beliefs in the second instance of the (non dynamic) Bayesian network, corresponding to the current levels of the competences, are propagated to the \emph{Term(inal) Conditions}, and recovered from there.
	\end{enumerate}
	\item The final state of the \emph{Term(inal) Conditions} correspond to the beliefs on the new levels of the competences, at $t=k$, and are stored in replacement of the beliefs corresponding to $t = 0$.
\end{enumerate}

	\subsection{Conditional probability distributions}
\label{sec:conditionals}

\noindent A common approach nowadays is to learn the conditional probability distributions from a large amount of data \cite{murphyDynamicBayesianNetworks2002,kaserDynamicBayesianNetworks2017}, which in this case would be about competence assessments by teachers. Unfortunately, in our case data is not quite abundant, particularly given the large amount of competences that underlie educational programmes, as well as the difference in perspectives that lead to different definitions, and hence competences maps, even for quite similar competences. Additionally, the competence-based educational model for high school education was introduced fifteen years \cite{secretariadeeducacionpublicaACUERDONumero4442008} ago and, despite the fact that many teachers were trained for competence-based teaching and evaluation, it is still work in progress, so data may be too ``dirty''.

So, instead of learning the conditional probability distributions from data, we decide to construct them from first principles on the basis of the different kinds of relationships in competences maps \cite{morales-gamboaProbabilisticRelationalLearner2017} (\autoref{tab:spegen} and \autoref{tab:subsup}), plus those added in its translations to a DBN (\autoref{tab:compevi} and \autoref{tab:paspre}), and a selection of numerical values for the fuzzy terms (\autoref{tab:numval})---in each table, the top row includes the possible values for the parent node, while the left column includes the possible values for the child node. Also, on the basis of a social constructivist perspective \cite{vygotskyMindSocietyDevelopment1978}, we decided to move away from the typical binary variables and to have three possible values for competence development (\emph{Low}, \emph{Medium}, and \emph{High}) representing no development (the student cannot perform the associated activity, even with scaffolding), partial development (the student can perform the associated activity, but only with scaffolding), and full development (the student can perform the associated activity on their own).

In the case of the inclusion/part-of relationship, in \autoref{tab:subsup}, the reasoning behind its design goes on the line that if someone cannot perform the super-competence (level Low), they have to get stuck in at least one sub-competence, which could, or could not, be a given sub-competence. So, from all possible configurations of competence levels among the sub-competences ($3^n$), we have to discard the cases in which level Low does not show ($2^n$). Then, we consider all cases in which a given sub-competence as level Low (as the other $n-1$ can have any value, they are $3^{n-1}$), as well as all cases in which the given sub-competence as other level, Medium or High ($3^{n-1} - 2^{n-1}$). Similarly, if someone can perform the super-competence but only with scaffolding, they cannot get stuck in any sub-competence but they would need help in at least one sub-competence. So, from all possible configurations of competence levels among the subcompetences ($2^n -1$, because we need to eliminate the case of all sub-competences to have level High), we consider the cases in which a given sub-competence has level Medium ($2^{n-1}$, as the others can any level other than Low), or High ($2^{n-1} - 1$, because not all others can have a High level). However, we have decided not to have a zero probability on any case, se we have adjusted the formulas in this case to allow some probability for level Low among sub-competences. Finally, if someone can perform the super-competence without any help, then they cannot get stuck nor need help on any sub-competence, so the probability for any sub-competence to have a level other than High should be zero, but we have opted for allowing small, non-zero probabilities for the other levels—more details of the rationale behind the design of all conditional probability distributions can be found in \cite{morales-gamboaProbabilisticRelationalLearner2017}

Concerning the design of the DBN, the final decision was on which numerical values to ascribe to the fuzzy terms used to describe the conditional probability distributions (transition probabilities). In this case, given the speed of decay of some probabilities in the conditional probability distribution for the inclusion/part-of relationship, we decided to define them on the bases of the standarized cummulative normal distribution so that the numerical values are those shown in \autoref{tab:numval}.

\begin{table}[!t]
	\renewcommand{\arraystretch}{1.3}
	\caption{Conditional probability distribution for the specialization/generalization relationship. Source \cite{morales-gamboaProbabilisticRelationalLearner2017}.}	
	\centering
	\begin{tabular*}{3.5in}{@{\extracolsep{\fill} } lccc}
		\hline\noalign{\smallskip}
		&\textbf{Low}&\textbf{Medium}&\textbf{High}\\
		\hline
		\noalign{\smallskip}
		\textbf{Low}&Large&Medium&Small\\
		\textbf{Medium}&Small&Large&Large\\
		\textbf{High}&Very small&Small&Medium\\
		\hline
	\end{tabular*}
	\label{tab:spegen}
\end{table}
\begin{table}[!t]
	\renewcommand{\arraystretch}{1.3}
	\caption{Conditional probability distribution for the inclusion/part-of relationship. The variable \(n\) stands for the number of sub-competences. Source \cite{morales-gamboaProbabilisticRelationalLearner2017}.}
	\centering
	\begin{tabular*}{3.5in}{@{\extracolsep{\fill} } lccc}
		\hline\noalign{\smallskip}
		&\textbf{Low}&\textbf{Medium}&\textbf{High}\\
		\hline
		\noalign{\smallskip}
		\textbf{Low}&$\displaystyle\frac{3^{n-1}}{3^n - 2^n}$&$\frac{1}{2^n}$&Very small\\
		\textbf{Medium}&$\displaystyle\frac{3^{n-1} - 2^{n-1}}{3^n - 2^n}$&$\frac{1}{2}$&Small\\
		\textbf{High}&$\displaystyle\frac{3^{n-1} - 2^{n-1}}{3^n - 2^n}$&$\frac{2^{n-1}-1}{2^n}$&Large\\
		\hline
	\end{tabular*}
	\label{tab:subsup}
\end{table}
\begin{table}[!t]
	\renewcommand{\arraystretch}{1.3}
	\caption{Conditional probability distribution for competence/evidence relationships.}	
	\centering
	\begin{tabular*}{3.5in}{@{\extracolsep{\fill} } lccc}
		\hline\noalign{\smallskip}
		&\textbf{Low}&\textbf{Medium}&\textbf{High}\\
		\hline
		\noalign{\smallskip}
		\textbf{Low}&Large&Medium&Small\\
		\textbf{Medium}&Small&Large&Medium\\
		\textbf{High}&Very small&Medium&Large\\
		\hline
	\end{tabular*}
	\label{tab:compevi}
\end{table}
\begin{table}[!t]
	\renewcommand{\arraystretch}{1.3}
	\caption{Conditional probability distribution for the past/present relationship.}	
	\centering
	\begin{tabular*}{3.5in}{@{\extracolsep{\fill} } lccc}
		\hline\noalign{\smallskip}
		&\textbf{Low}&\textbf{Medium}&\textbf{High}\\
		\hline
		\noalign{\smallskip}
		\textbf{Low}&Large&Very small&Tiny\\
		\textbf{Medium}&Very small&Large&Very small\\
		\textbf{High}&Tiny&Tiny&Large\\
		\hline
	\end{tabular*}
	\label{tab:paspre}
\end{table}
\begin{table}[!t]
	\renewcommand{\arraystretch}{1.3}
	\caption{Numerical values associated to the fuzzy terms used to describe the conditional probability distributions.}	
	\centering
	\begin{tabular*}{3.5in}{@{\extracolsep{\fill} } lrl}
		\hline\noalign{\smallskip}
		\textbf{Term}&$\sigma$&\textbf{Value}\\
		\hline
		\noalign{\smallskip}
		{Large}&0&0.5\\
		{Medium}&-1&0.15865525393145707\\
		{Small}&-2&0.02275013194817919\\
		{Very small}&-3&0.00134989803163009\\
		{Tiny}&-4&0.00003167124183311\\
		\hline
	\end{tabular*}
	\label{tab:numval}
\end{table}

\section{Models of prototypical students}
\label{sec:models}

\noindent In order to observe the behaviour of student models created in such a way, we simulate a course devoted to the development of the specialized competences included in the map shown in \autoref{fig:projectCollaborationMap}, and students exhibiting three prototypical performances: low to medium performance, medium to high performance but with final failure, and two terms medium to high performance (second course is assumed not devoted to the development of such competences, but some activities make use of them). The courses are supposed to run for fourteen weeks (each one corresponding to a time slice), plus two weeks for revisions and additional examinations, and five weeks of holidays before the start of the next term (\autoref{tab:protstud}).

\begin{table}[!t]
	\renewcommand{\arraystretch}{1.3}
	\caption{Evidences for simulated students: low to medium performance (L2M), medium to high performance (M2H), but with final product missing, and two terms medium to high performance (LT M2H)). All competences are related to a project, and numeric values (0, 1, and 2) are assigned to competence levels (\emph{Low}, \emph{Medium}, and \emph{High}, respectively).}	
	\centering
	\begin{tabular*}{3.5in}{@{\extracolsep{\fill} } rcccc}
		\hline\noalign{\smallskip}
		\textbf{Week}&Competence&\textbf{L2H}&\textbf{M2H}&\textbf{LT M2H}\\
		\noalign{\smallskip}
		\hline
		1&Propose&0&1&1\\
		2&Contribute&0&1&1\\
		4&Propose&1&2&2\\
		5&Contribute&0&1&1\\
		7&Collaborate&0&1&1\\
		10&Propose&1&2&2\\
		11&Contribute&1&2&2\\
		14&Collaborate&1&0&2\\
		23&Propose&&&2\\
		24&Contribute&&&2\\
		25&Collaborate&&&2\\
		35&Collaborate&&&2\\
		\hline
	\end{tabular*}
	\label{tab:protstud}
\end{table}

\begin{figure*}[!t]
	\centering
	\subfloat[Average]{\includegraphics[width=2.5in]{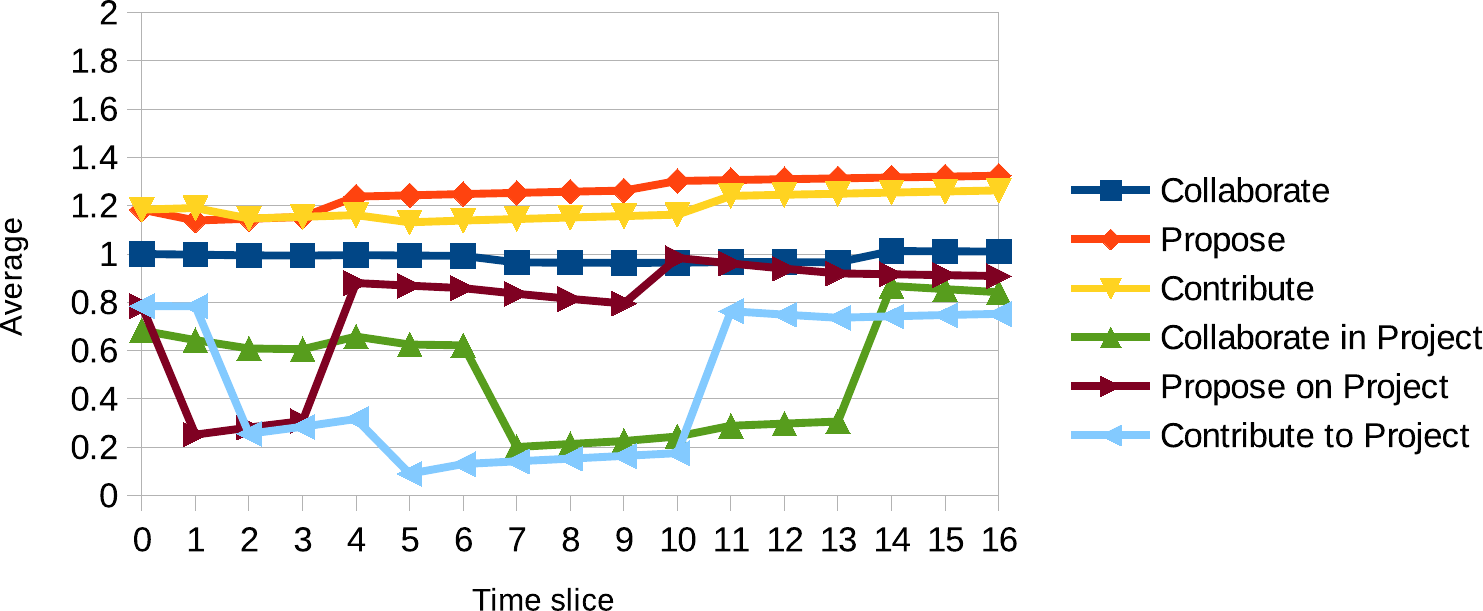}%
	\label{fig:l2m16-average}}
	\hfil
	\subfloat[Uncertainty]{\includegraphics[width=2.5in]{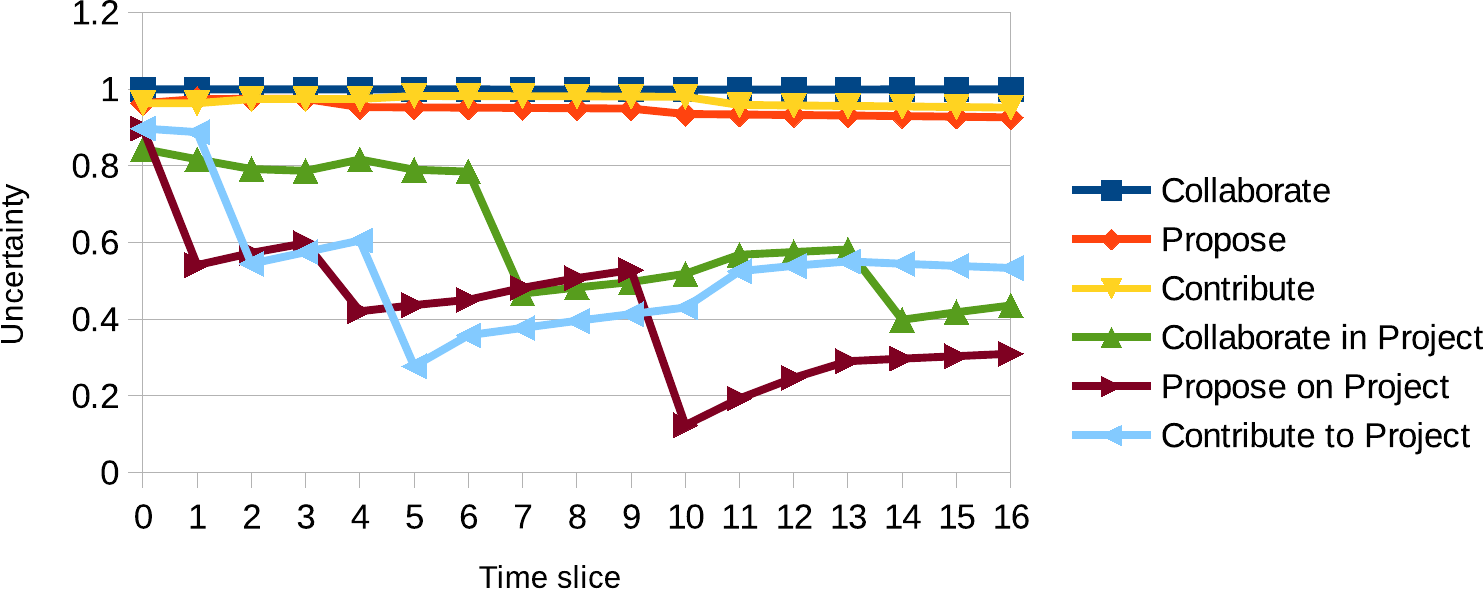}%
	\label{fig:l2m16-uncertainty}}
	\caption{Evolving beliefs on competence levels of the student with low to medium performance. Low, Medium, and High competence levels are translated to numbers (0, 1, and 2, respectively) and then beliefs are represented by both the average (\autoref{eqn:average}) and uncertainty of their probability distributions, the later calculated as normalized entropy \cite{sucarProbabilisticGraphicalModels2015a} (\autoref{eqn:uncertainty}).}
	\label{fig:l2m16}
\end{figure*}
\begin{figure*}[!t]
	\centering
	\subfloat[Average]{\includegraphics[width=2.5in]{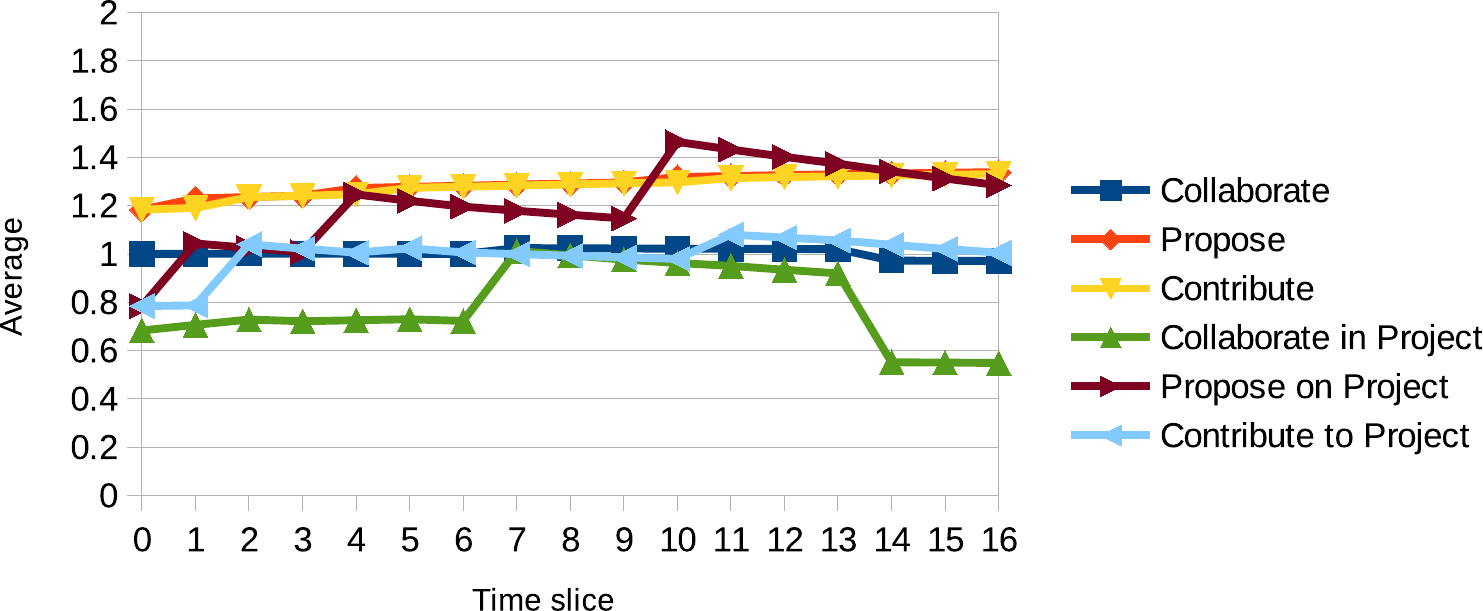}%
	\label{fig:m2h16-average}}
	\hfil
	\subfloat[Uncertainty]{\includegraphics[width=2.5in]{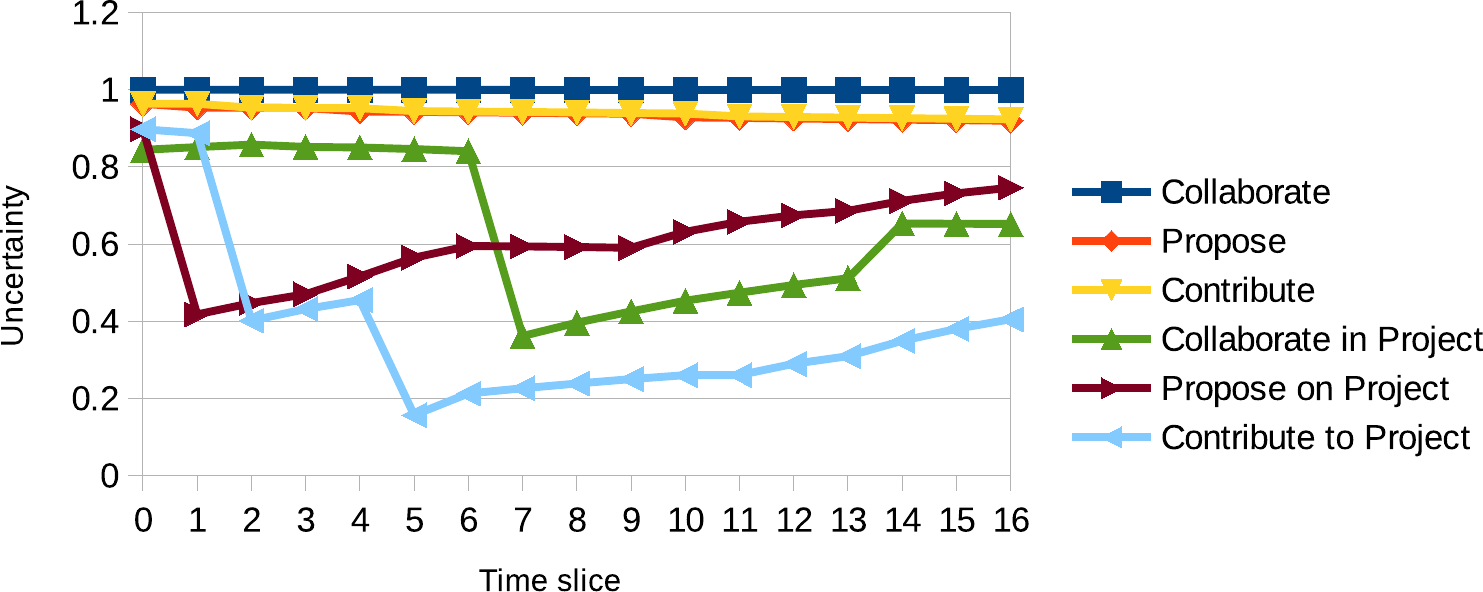}%
		\label{fig:m2h16-uncertainty}}
	\caption{Evolving beliefs (average and uncertainty) on competence levels of the student with medium to high performance, who failed on the last evaluation.}
	\label{fig:m2h16}
\end{figure*}
\begin{figure*}[!t]
	\centering
	\subfloat[Average]{\includegraphics[width=2.5in]{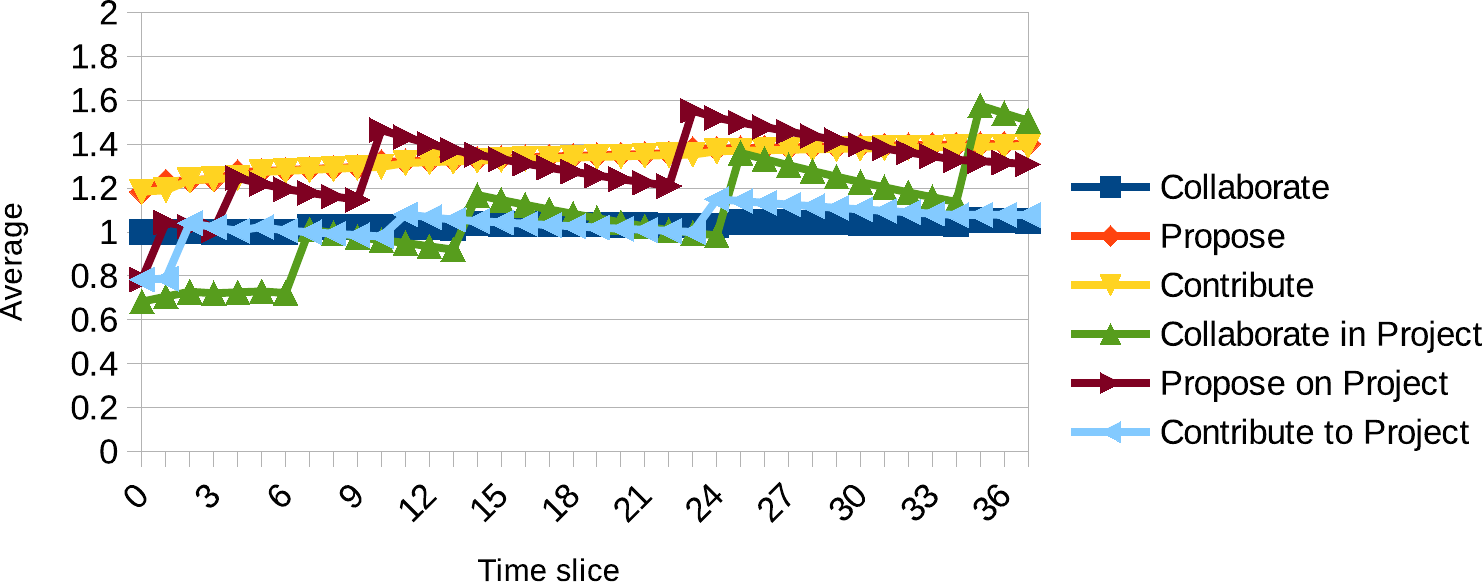}%
		\label{fig:m2h37-average}}
	\hfil
	\subfloat[Uncertainty]{\includegraphics[width=2.5in]{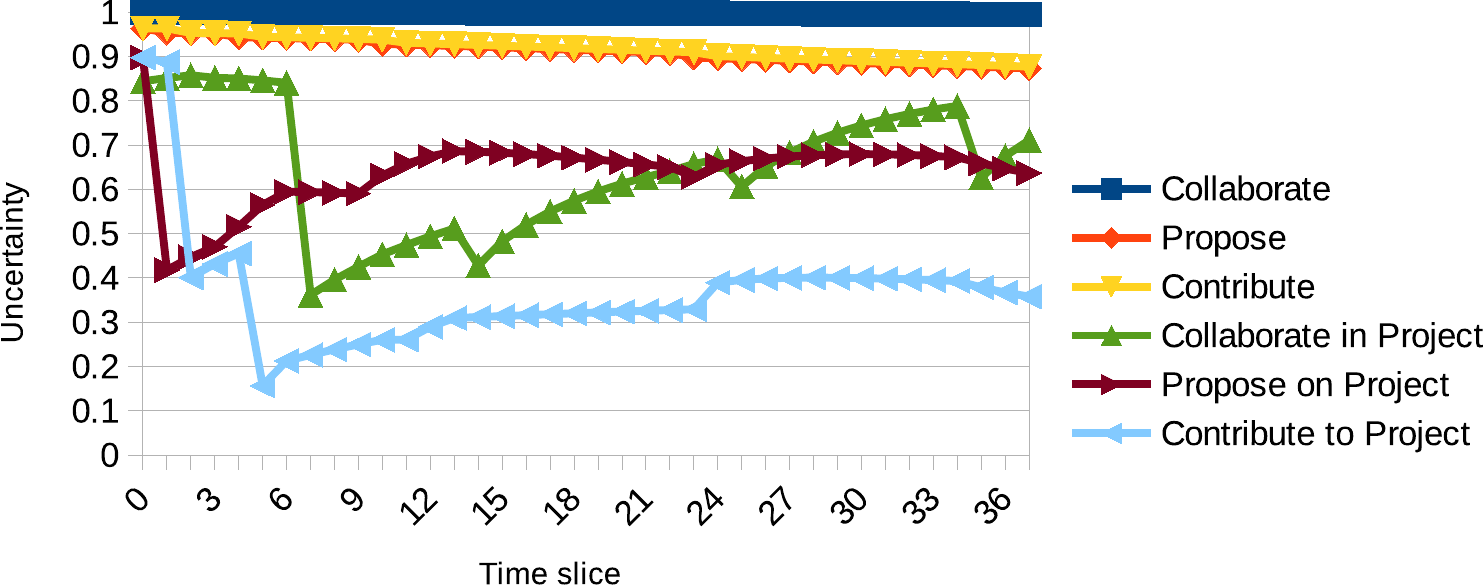}%
		\label{fig:m2h37-uncertainty}}
	\caption{Evolving beliefs (average and uncertainty) on competence levels of the student with medium to high performance along two periods.}
	\label{fig:m2h37}
\end{figure*}

Figures \autoref{fig:l2m16-average} and \autoref{fig:l2m16-uncertainty} show the evolving beliefs on the competence levels of the student with low to medium performance. Low, Medium, and High competence levels are translated to numbers ($i = 0$, 1, and 2, respectively) and then beliefs are represented by both the average,
\begin{equation}
	\sum p(i)i,
	\label{eqn:average}
\end{equation}
and uncertainty of their probability distributions, the later calculated as normalized entropy \cite{sucarProbabilisticGraphicalModels2015a},
\begin{equation}
	\frac{\sum{p(i)\ln(p(i))}}{\ln(\frac{1}{3})}.
	\label{eqn:uncertainty}
\end{equation}
Figures \autoref{fig:m2h16-average} and \autoref{fig:m2h16-uncertainty} show the case for the student with medium to high performance with final failure, while figures \autoref{fig:m2h37-average} and \autoref{fig:m2h37-uncertainty} show the case for the student with medium to high performance across two terms.

In all cases, uncertainty increases in time slices with no evidence, and it usually decreases on the presence of evidence. However, in the case of the student exhibiting low to medium performance, after two evidences for low level of competence at contributing to a project, an evidence of medium level of competence at time slice 11 increases uncertainty. A similar behaviour can be observed when evidence of low level performance arrives in time slice 14 for the student with medium to high performance: many beliefs go down but their uncertainties increase.

The prototype is more reluctant to accept evidence of good performance than of low or medium performance, due to the asymmetry in the conditional probability distribution in \autoref{tab:compevi}. Yet, after accumulating evidences from two terms in the case of a student with high performance, the beliefs show a tendency to slowly go up.	It is clear in all cases that beliefs on the development of the more general competences move much slowly, as there is no direct evidence of their level. Yet beliefs show a tendency to go up, and decrease their uncertainty, suggesting some positive development is going on after all.

\section{Comparison}
\label{sec:comparison}

\noindent In order to evaluate the results obtained so far, and the overall design of our system, we decided to compare the estimations it produces against estimations provided by teachers. We designed a questionnaire in which we ask teachers to estimate the levels of the more specific competences of our prototypical students (given the evidence in time slices 3, 6, 7, 12, 14, and 35) and to provide a degree of certainty for their estimations. Finally, we ask them to estimate the levels of the more general competences at the end of the course, and their degree of certainty on that as well. We used the questionnaire to run an online poll among colleagues and doctoral students, 20 of which answered it---from a population of circa three hundred, so we can claim significance of results only at the level of the sample. Among the participants, 65\%  said they are full time teachers, 25\% said they are subject teachers, 5\% said they are mainly researchers, and another 5\% said they do not teach. 60\% of the participants said they teach mainly at undergraduate level, 35\% said they are mainly postgraduate teachers, and only 5\% classify themselves mainly as high school teachers.

The teachers provided their estimations for the competence levels using a Likert scale with five values: (0) Low, (0.5) Rather Low, (1) Medium, (1.5) Rather High, and (2) High. They provided their degrees of certainty in a similar scale, which we show here inverted and normalized to describe their degree of uncertainty. We ran the Shapiro-Wilk test of normality on the estimations of competence levels provided by the teachers, and we found that most distributions are far from normal, so we proceed to the analysis of results using non parametric methods. For each question (58 in total) we calculated the first, second (median), and third quartiles of competence levels estimated, or degree of uncertainty. In order to measure the consistency among the answers using the interquartile range (IQR), we calculated IQR per question, and then we calculated the same quartiles for these new (meta) data. The results, presented in \autoref{tab:consistency}, show a tendency toward the minimum distance, or less, among the responses provided by the teachers, suggesting a high consistency among them.

\begin{table}[!t]
	\renewcommand{\arraystretch}{1.3}
	\caption{Non parametric consistency analysis measuring the Interquartile Range (IQR) among responses, per question, and then calculating the quartiles and the new IQR among the IQRs previously calculated.}	
	\centering
	\begin{tabular*}{2.5in}{@{\extracolsep{\fill} } lrr}
		\hline\noalign{\smallskip}
		\textbf{Quartile}&\textbf{Average}&\textbf{Uncertainty}\\
		\hline
		\noalign{\smallskip}
		{Maximum}& 1.0 & 0.5\\
		{Third} & 0.5   & 0.25\\
		{Second (Median)}  & 0.5   & 0.25\\
		{First} & 0.125 & 0.25\\
		{Minimum} & 0.0 & 0.0 \\
		\noalign{\smallskip}
		\hline
		\noalign{\smallskip}
		\textbf{IQR}     & \textbf{0.375} & \textbf{0.0}\\
		\noalign{\smallskip}
		\hline
	\end{tabular*}
	\label{tab:consistency}
\end{table}

\begin{figure*}[!t]
	\centering
	\subfloat[Average]{\includegraphics[width=2.5in]{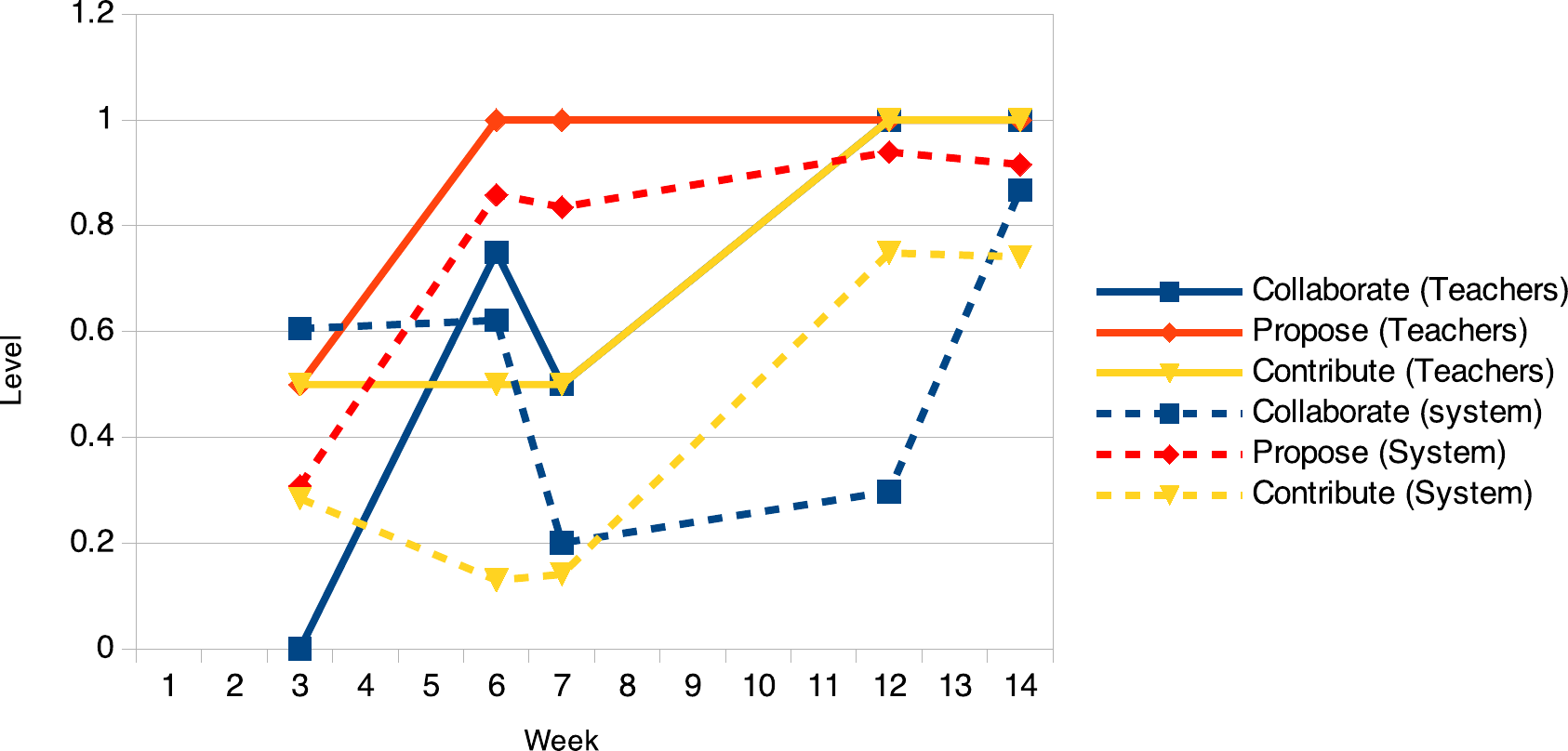}%
		\label{fig:l2m16-average-comparison}}
	\hfil
	\subfloat[Uncertainty]{\includegraphics[width=2.5in]{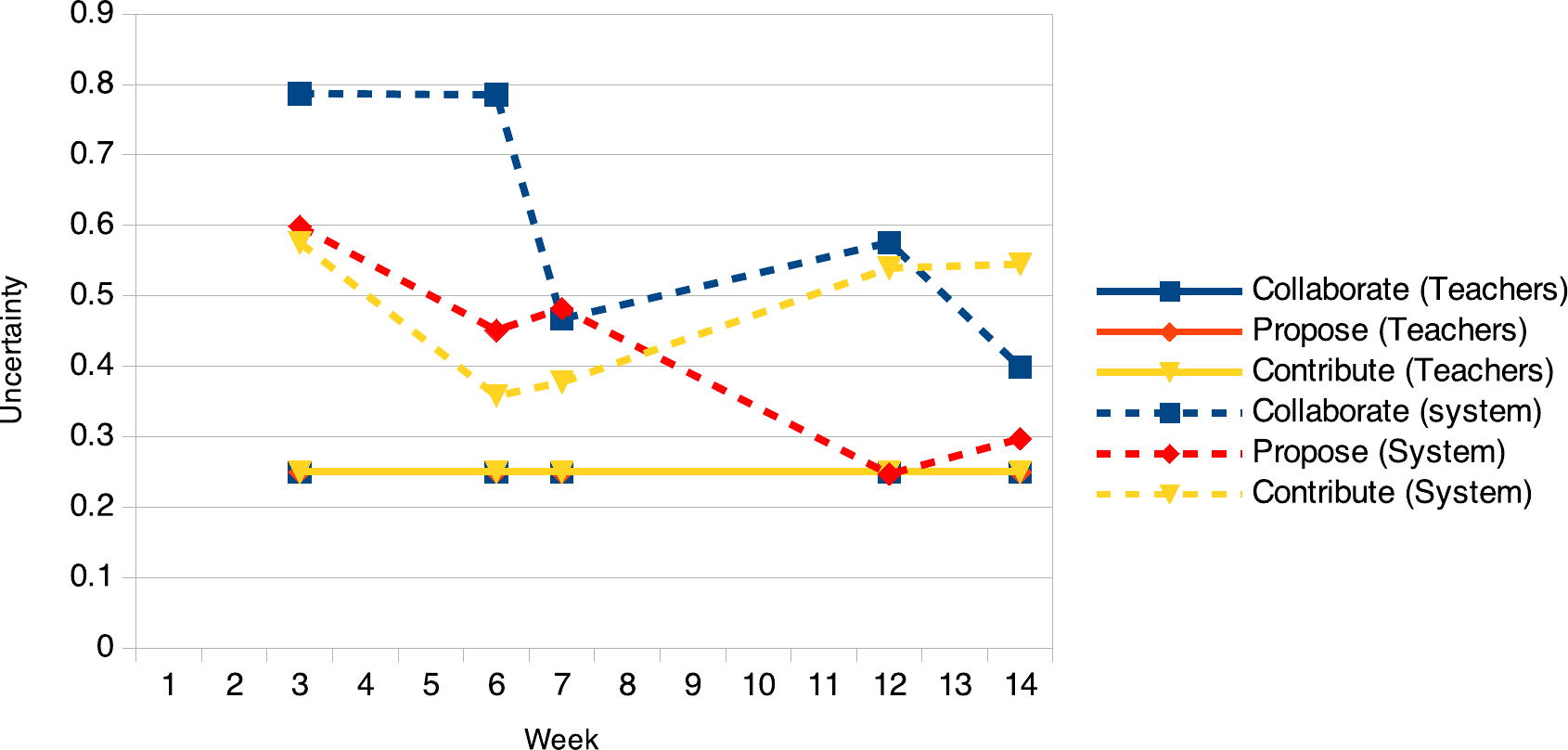}%
		\label{fig:l2m16-uncertainty-comparison}}
	\caption{Comparison of estimates by teachers and the system of competence levels of the student with low to medium performance.}
	\label{fig:l2m16-comparison}
\end{figure*}

\begin{figure*}[!t]
	\centering
	\subfloat[Average]{\includegraphics[width=2.5in]{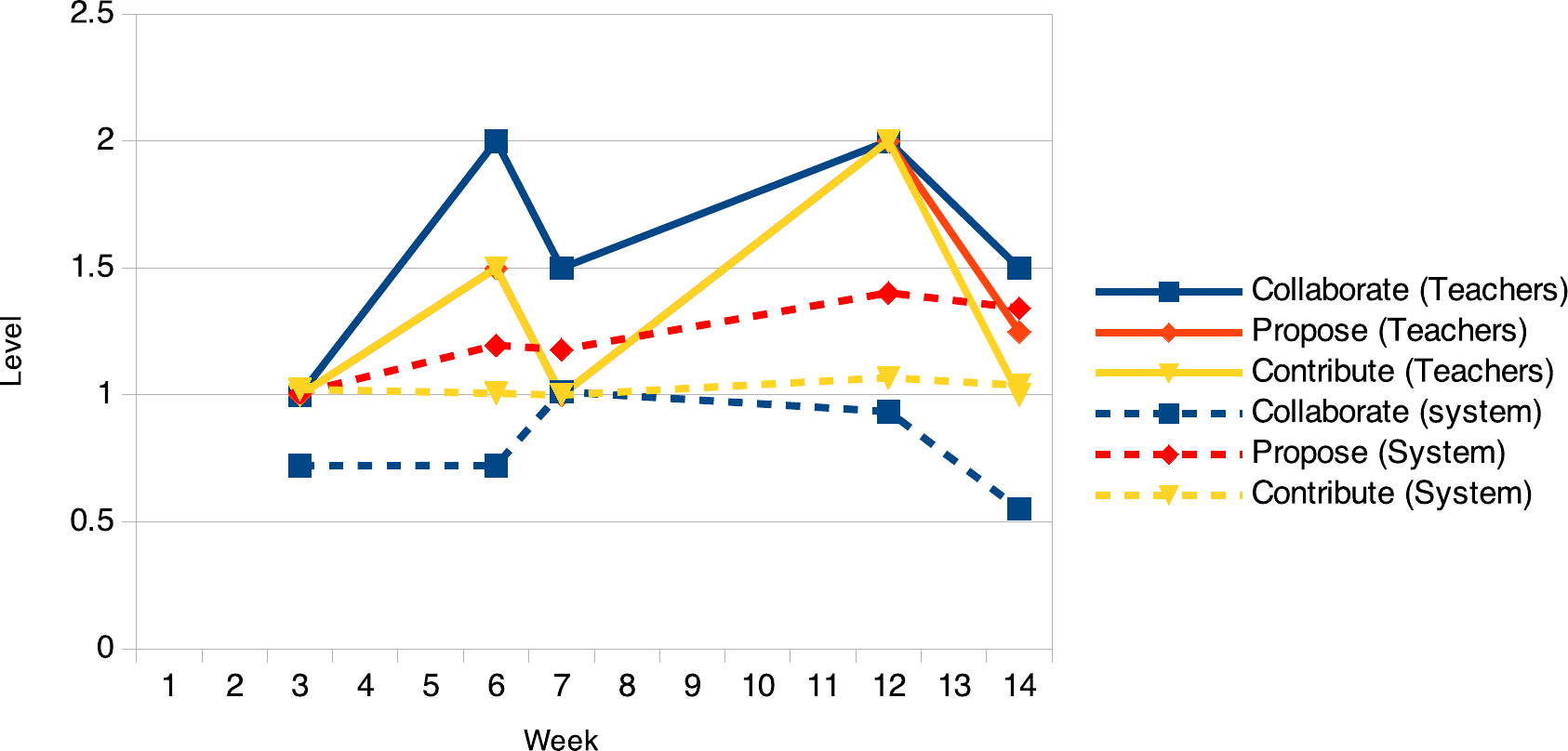}%
		\label{fig:m2h16-average-comparison}}
	\hfil
	\subfloat[Uncertainty]{\includegraphics[width=2.5in]{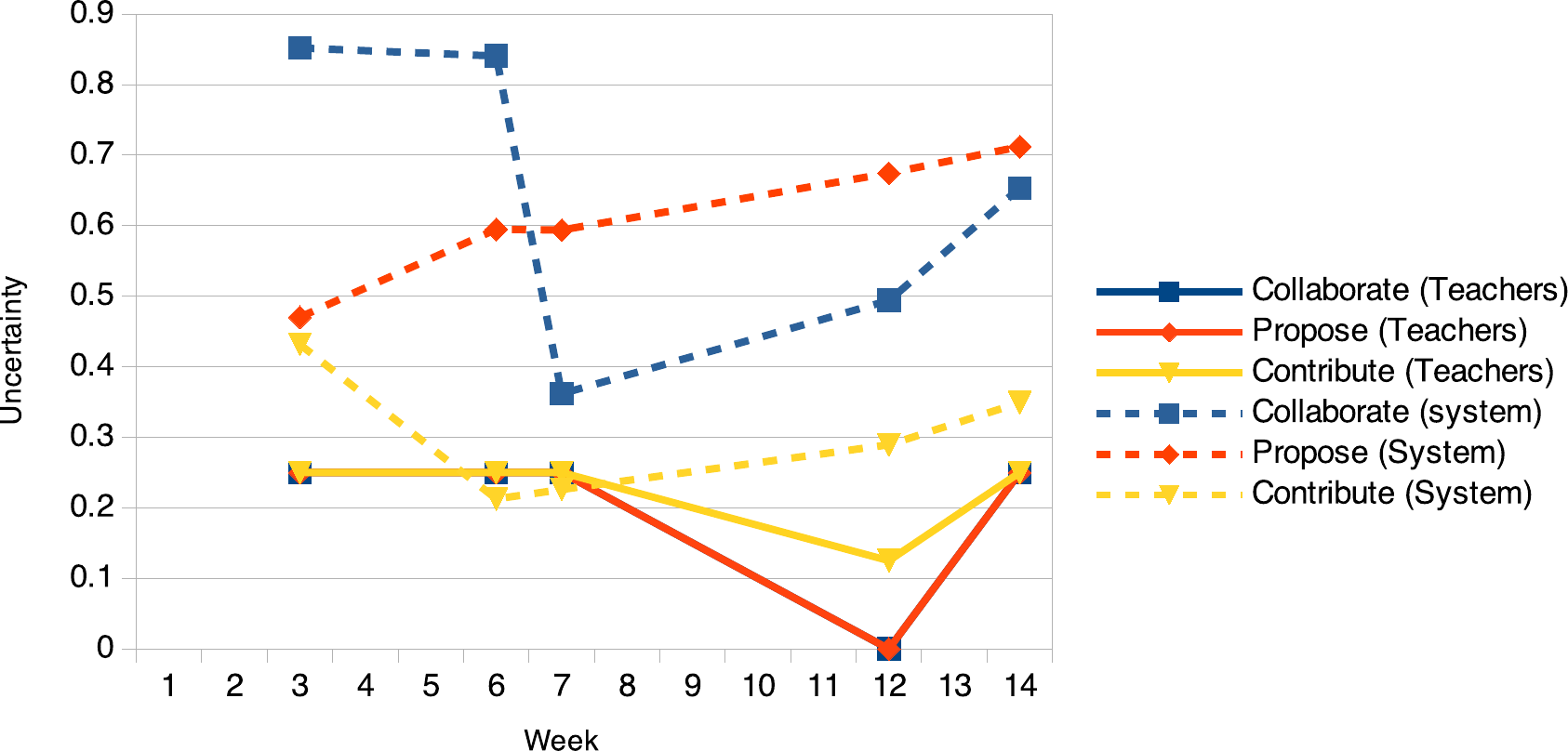}%
		\label{fig:m2h16-uncertainty-comparison}}
	\caption{Comparison of estimates by teachers and the system of competence levels of the student with medium to high performance, who failed on the last evaluation.}
	\label{fig:m2h16-comparison}
\end{figure*}

\begin{figure*}[!t]
	\centering
	\subfloat[Average]{\includegraphics[width=2.5in]{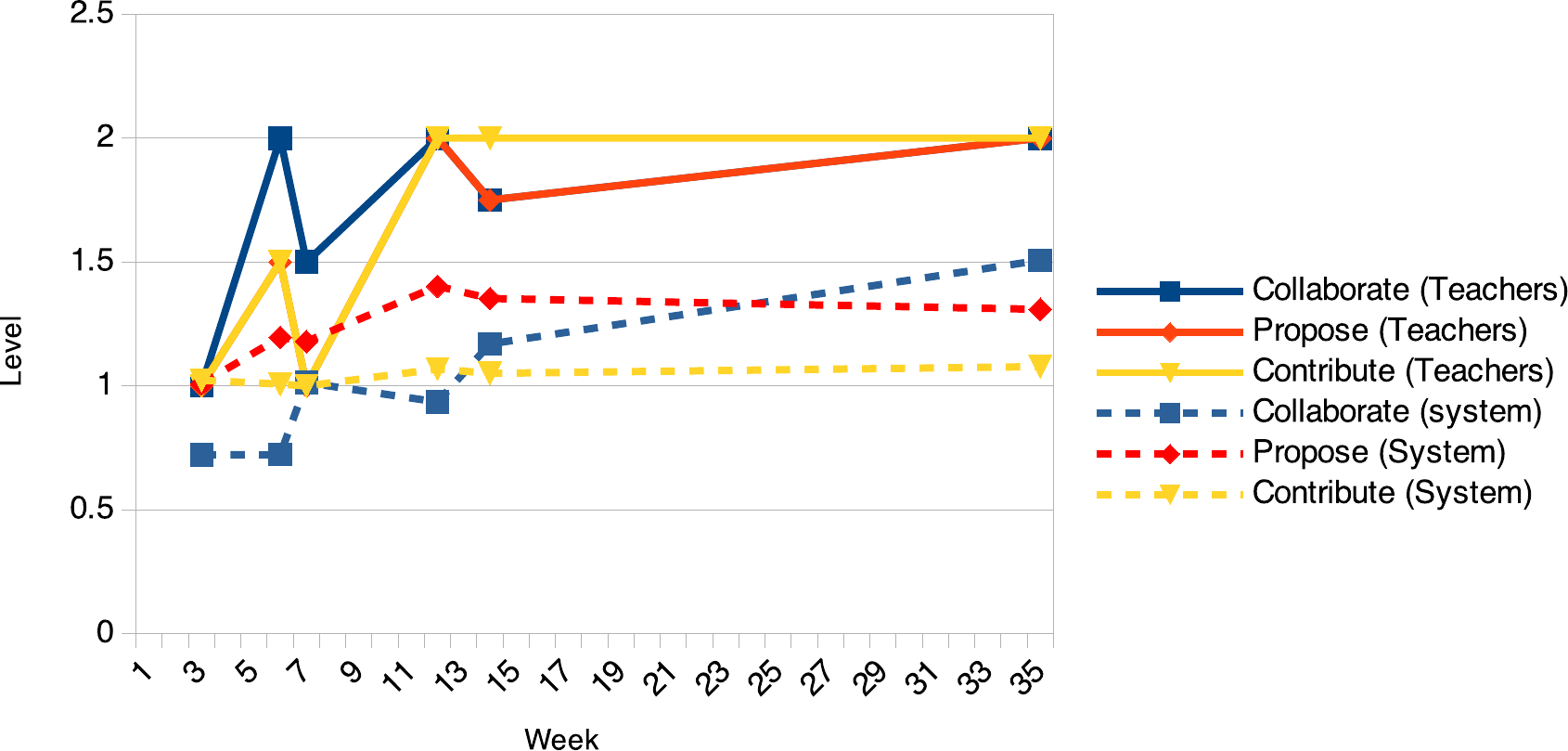}%
		\label{fig:m2h37-average-comparison}}
	\hfil
	\subfloat[Uncertainty]{\includegraphics[width=2.5in]{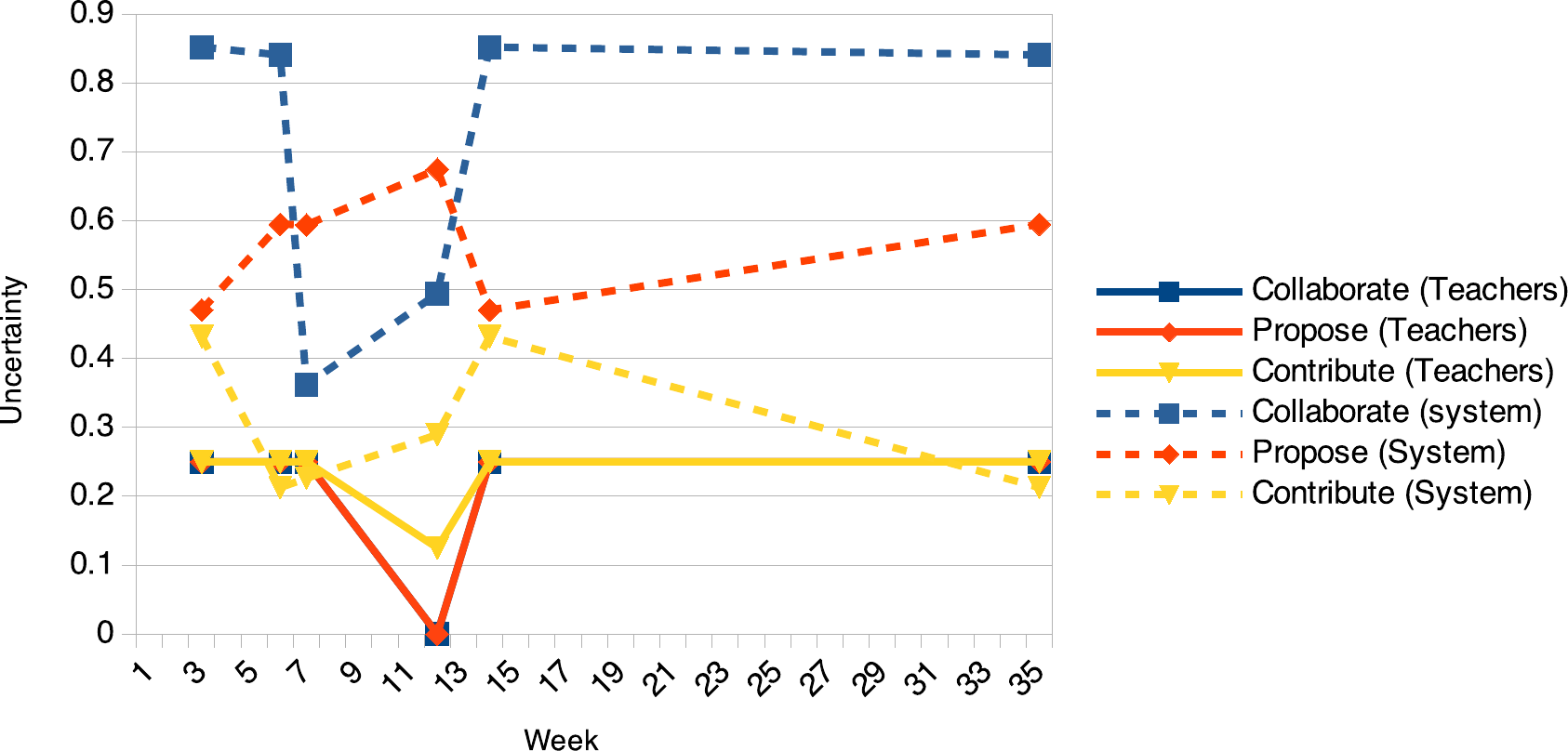}%
		\label{fig:m2h37-uncertainty-comparison}}
	\caption{Comparison of estimates by teachers and the system of competence levels of the student with medium to high performance along tow periods.}
	\label{fig:m2h37-comparison}
\end{figure*}

Then we compared the responses provided by the teachers against the estimations provided by the system, first graphically as in \autoref{fig:l2m16-comparison}, \autoref{fig:m2h16-comparison}, and \autoref{fig:m2h37-comparison}. We noticed that although the estimations provided by the system are, in general, lower than the median of those provided by the teachers (figures \autoref{fig:l2m16-average-comparison}, \autoref{fig:m2h16-average-comparison}, and \autoref{fig:m2h37-average-comparison}), there seems to be a correlation between them, so we calculated their Pearson's coefficient of correlation, $r$, and the results are shown in \autoref{tab:pearson}. They suggest that the teachers and the system followed the same pattern in estimating competence levels when there was more or less frequent evidence for the competence, and they indicate an unsurprising behaviour, particularly in the case of low to medium performance. The differences grow when there are surprises in performance (e.g. the student shows steady improvment but the fails at one examination), and when there is fewer evidence, as is the case for the competence of collaborating in project. Then we calculated first and third quartiles among the estimations provided by teachers, and we compared them against the estimations generated by the system. The results were that, in general, the estimations provided by the system are out of the range defined by the first and third quartile. In the case of the low to medium performance student, only 40\% of system estimations are in the range, whereas in the other two cases the percentages were only 26.7\% (low to medium with final failure) and 22.2\% (medium to high across two terms). So, these results confirm the significance of difference in estimations by the teachers and the system shown in \autoref{fig:l2m16-comparison}, \autoref{fig:m2h16-comparison}, and \autoref{fig:m2h37-comparison}. Furthermore, it shows the difference is bigger in the cases of evidence of medium to high performance.


Regarding the degree of uncertainty in the estimations, uncertainty in the beliefs maintained by the system shows relatively high sensitivity to changes in the evidence, whereas the uncertainty in estimations by the teachers seems almost stable. Finally, the comparison of estimations of the final state of development of the most generic competences, and the corresponding uncertainties, by the teachers and the system, shown in \autoref{tab:general-competences}, suggest a similar pattern of the teachers being more optimistic by providing higher estimations of competence development, together with much lower levels of uncertainty.

We have carried out some attempts to get the system estimations  closer to those provided by the teachers. That is to say, we have tried to  relax the conditional probability distributions for the relationship between previous and current state of a node in the dynamic Bayesian network (\autoref{tab:paspre}), as a way of acknowledging student state changes along their studies. Unfortunately, those adjustments led to quicker increases in uncertainty along weeks, which seemed less realistic than the ones shown in this paper.

\begin{table*}[!t]
	\caption{Pearson correlation coefficientes between competence level estimations by the teachers and the system for students of low to medium performance (L2M), medium to high but failing on final evaluation (M2H), and long term steady medium to high (LT M2H).}	
	\centering
	\begin{tabular*}{5.0in}{@{\extracolsep{\fill} } lccc}
		\hline\noalign{\smallskip}
		&  \multicolumn{3}{c}{\textbf{Competence level}}\\ 
		\hline
		\noalign{\smallskip}
		&\parbox{3cm}{\centering\textbf{Collaborate in project}}&
		\parbox{2cm}{\centering\textbf{Propose on project}}&
		\parbox{2cm}{\centering\textbf{Contribute to project}}\\
		\noalign{\smallskip}
		\hline
		\noalign{\smallskip}
		\textbf{L2M}&0.1026&0.9871&0.9808\\
		\textbf{M2H}&0.2385&0.7406&0.6307\\
		\textbf{LT M2H}&0.3458&0.8159&0.8512\\
		\hline
	\end{tabular*} 
	\label{tab:pearson}
\end{table*}

\begin{table*}[!t]
	\caption{Comparison of estimates and incertainties as provided by the teachers and the system regarding the development of the most generic competences—to collaborate (Col), to propose (Prop), and to contribute (Cont)—by students of low to medium performance (L2M), medium to high performance with failure in final evaluation (M2H), and steady medium to high performance along two terms (LT M2H).}	
	\centering
	\begin{tabular*}{\textwidth}{@{\extracolsep{\fill}}l|ccc|ccc|ccc}
		\hline\noalign{\smallskip}
		&  \multicolumn{3}{c}{\textbf{L2M}}&\multicolumn{3}{c}{\textbf{M2H}}&
		\multicolumn{3}{c}{\textbf{LT M2H}}\\ 
		\noalign{\smallskip}
		\hline
		\noalign{\smallskip}
		&\parbox{0.7cm}{\centering\textbf{Col}}&
		\parbox{0.9cm}{\centering\textbf{Prop}}&
		\parbox{0.9cm}{\centering\textbf{Cont}}&
		\parbox{0.7cm}{\centering\textbf{Col}}&
		\parbox{0.9cm}{\centering\textbf{Prop}}&
		\parbox{0.9cm}{\centering\textbf{Cont}}&
		\parbox{0.7cm}{\centering\textbf{Col}}&
		\parbox{0.9cm}{\centering\textbf{Prop}}&
		\parbox{0.9cm}{\centering\textbf{Cont}}\\
		\noalign{\smallskip}
		\hline
		\noalign{\smallskip}
		\parbox{2.05cm}{\begin{flushleft}\textbf{Teacher estimation (median)}\end{flushleft}}&1&1&1&1.5&1.5&1.25&2&2&1.75\\
		\parbox{2.05cm}{\begin{flushleft}\textbf{Teacher uncertainty (median)}\end{flushleft}}&0.25&0.25&0.25&0.25&0.25&0.25&0.25&0.25&0.25\\
		\noalign{\smallskip}
		\hline
		\noalign{\smallskip}
		\parbox{2.05cm}{\begin{flushleft}\textbf{System estimation (median)}\end{flushleft}}&1.01&1.32&1.25&0.97&1.33&1.32&1.05&1.40&1.39\\
		\parbox{2.05cm}{\begin{flushleft}\textbf{System uncertainty}\end{flushleft}}&1&0.93&0.96&1&0.92&0.93&0.99&0.87&0.88\\
		\noalign{\smallskip}		
		\hline
	\end{tabular*} 
	\label{tab:general-competences}
\end{table*}

\section{Conclusions}
\label{sec:conclusions}

We have shown a way of creating overlay student models as dynamic Bayesian networks built on top of competences maps including generalization{/}specialization and inclusion/part-of relationships. It works by defining a conditional probability distribution per type of relationship (and cardinality, in the case of inclusion/part-of relationships), so it can be applied to any map restricted to those common relationships. This approach provides a general way for assigning weights to the relationships between competences, which does not make use of the fine grained composition of the competences to construct nor evaluate evidence, considering competences very much as holistic entities. In that sense, it is quite different from fine grained approaches typical in the field, but it would be much easier to make it work on real life conditions, provided it delivers reasonable results.

The results obtained from implementing this method on a given competences map, performing the modelling of competence development by some prototypical students, and then comparing the competence levels estimated by the prototype against those estimated by teachers, suggest that the inferencing carried out by the dynamic Bayesian network goes along what teachers estimate from  evidences of performance by students, but teachers seem to be significantly more optimistic and confident on competence development by students, particularly when they get evidence of good performances. Furthermore, they seem willing to generalize the evidence gathered in relation to more specific competences, so they estimate similar levels of development for the more generic ones, and they do so without loosing certainty in their estimations. In general, it seems teachers give more attention to current evidence than our system does, and are more relaxed in considering previous evidence too. Furthermore, they seem to assume the student is engaged in learning, so they expect an improvement in competence levels, and positive evidence works as a confirmation for their expectations. In contrast, our system assumes decaying of competence levels unless evidence on the contrary is provided; a closed world assumption the teachers do not hold.

We have assumed all evidences to be hard ones, from what we gathered as prototypical performances, but nevertheless invented ones, and we have no information regarding the expertise on competence estimation by the participants on our study. Furthermore, the task imposed on the participants is not what they are used to do: a kind of meta-analysis on the evidence produced by others, instead of directly observing the performance of students and genereting the evidence themselves (what it would be their actual task if the system gets implemented for some educational programme). So we acknowledge we are still far from fully evaluating the modelling of the development of competence levels by our system.

Future work includes the discussion of the implications of this evaluation, and a subsequent fine tuning of our conditional probability distributions. An explicit modelling of the estimations provided by teachers, by means of adjusting the conditional probability distributions, would also be useful to understanding their reasoning. There is also work to do in expanding the development and testing with evidence produced from real (historical) data by expert competence evaluators, and in comparing the inferences performed by the system against their estimations. Beyond that, we would expect to incorporate the full map of transversal competences for Mexican high school students, and to do real-time student modelling, considering not only information provided by teachers, but information gathered from other sources, inside and outside schools.

\section*{Acknowledgments}

This work has been partially funded by the Common Space for Distance Higher Education (ECOESAD) and the Program for Teachers Professional Development (PRODEP). The core of our implementations is based on the SMILE reasoning engine for graphical probabilistic models, while images of them included in this paper were created using the GeNIe Modeler, both available free of charge for academic research and teaching use from \href{http://www.bayesfusion.com/}{BayesFusion, LLC}. We thanks colleagues and postgraduate students that voluntarily participated in our poll.




\bibliographystyle{IEEEtran}
\bibliography{IEEEabrv,dbn-sm}

\begin{thebibliography}{10}
\providecommand{\url}[1]{#1}
\csname url@samestyle\endcsname
\providecommand{\newblock}{\relax}
\providecommand{\bibinfo}[2]{#2}
\providecommand{\BIBentrySTDinterwordspacing}{\spaceskip=0pt\relax}
\providecommand{\BIBentryALTinterwordstretchfactor}{4}
\providecommand{\BIBentryALTinterwordspacing}{\spaceskip=\fontdimen2\font plus
\BIBentryALTinterwordstretchfactor\fontdimen3\font minus
  \fontdimen4\font\relax}
\providecommand{\BIBforeignlanguage}[2]{{%
\expandafter\ifx\csname l@#1\endcsname\relax
\typeout{** WARNING: IEEEtran.bst: No hyphenation pattern has been}%
\typeout{** loaded for the language `#1'. Using the pattern for}%
\typeout{** the default language instead.}%
\else
\language=\csname l@#1\endcsname
\fi
#2}}
\providecommand{\BIBdecl}{\relax}
\BIBdecl

\bibitem{gordonKeyCompetencesEurope2009}
J.~Gordon, G.~Halasz, M.~Krawczyk, T.~Leney, A.~Michel, D.~Pepper,
  E.~Putkiewicz, and J.~Wi{\'s}niewski, ``\BIBforeignlanguage{en}{Key
  {{Competences}} in {{Europe}}: {{Opening Doors For Lifelong Learners Across}}
  the {{School Curriculum}} and {{Teachers Education}}},'' {CASE \textendash{}
  Center for Social and Economic Research}, {Warsaw, Poland}, Tech. Rep.
  87/2009, 2009.

\bibitem{lurieDeconstructingCompetencybasedEducation2017}
H.~Lurie and R.~Garrett, ``Deconstructing competency-based education: {{An}}
  assessment of institutional activity, goals, and challenges in higher
  education,'' \emph{The Journal of Competence-Based Education}, vol.~2, no.~3,
  pp. 1--19, 2017.

\bibitem{nunezcortesModeloCompetencialCompetencia2016}
J.~A. N{\'u}{\~n}ez~Cort{\'e}s, ``El modelo competencial y la competencia
  comunicativa en la educaci{\'o}n superior en {{Am{\'e}rica Latina}},''
  \emph{Foro de Educaci{\'o}n}, vol.~14, no.~20, pp. 467--488, 2016.

\bibitem{moralesIntelligentEnvironmentDistance2009}
R.~Morales, ``Towards an intelligent environment for distance learning,''
  \emph{World Journal on Educational Technology}, vol.~1, no.~2, pp. 110--117,
  2009.

\bibitem{morales-gamboaProbabilisticRelationalLearner2017}
R.~{Morales-Gamboa}, E.~{Sucar-Succar}, E.~{Ruiz-Hern{\'a}ndez}, M.~E.
  {Chan-N{\'u}{\~n}ez}, and S.~C. Gonz{\'a}lez~Flores,
  ``\BIBforeignlanguage{en}{Probabilistic relational learner models based on
  competence maps},'' \emph{\BIBforeignlanguage{en}{Research in Computing
  Science}}, vol. 146, pp. 77--86, 2017.

\bibitem{competency_data_working_group_1484.20.1-2007_2008}
{Competency Data Working Group}, \emph{\BIBforeignlanguage{en}{1484.20.1-2007
  {{IEEE Standard}} for {{Learning Technology}}\textemdash{{Data Model}} for
  {{Reusable Competency Definitions}}}}, ser. {{IEEE Standard}} for {{Learning
  Technology}}.\hskip 1em plus 0.5em minus 0.4em\relax {New York, NY}: {The
  Institute of Electrical and Electronics Engineers, Inc}, Jan. 2008.

\bibitem{imsgloballearningconsortiumIMSReusableDefinition2002}
I.~G.~L. Consortium, ``\BIBforeignlanguage{en}{{{IMS Reusable Definition}} of
  {{Competency}} or {{Educational Objective Specification}}},'' {IMS Global
  Learning Consortium, Inc.}, Tech. Rep. Version 1.0 Final Specification, 2002.

\bibitem{sampsonDevelopingCommonMetadata2007}
D.~Sampson, P.~Karampiperis, and D.~Fytros,
  ``\BIBforeignlanguage{en}{Developing a common metadata model for competencies
  description},'' \emph{\BIBforeignlanguage{en}{Interactive Learning
  Environments}}, vol.~15, no.~2, pp. 137--150, 2007.

\bibitem{paquetteLearningDesignBased2006}
G.~Paquette, M.~L{\'e}onard, K.~{Lundgren-Cayrol}, S.~Mihaila, and D.~Gareau,
  ``Learning {{Design}} based on {{Graphical Knowledge}}-{{Modelling}},''
  \emph{Educational Technology \& Society}, vol.~9, no.~1, pp. 97--112, 2006.

\bibitem{stoofWebbasedSupportConstructing2007}
A.~Stoof, R.~L. Martens, and J.~J.~G. van Merri{\"e}nboer,
  ``\BIBforeignlanguage{en}{Web-based support for constructing competence maps:
  Design and formative evaluation},'' \emph{\BIBforeignlanguage{en}{Educational
  Technology Research and Development}}, vol.~55, no.~4, pp. 347--368, 2007.

\bibitem{pepiotUECMLUnifiedEnterprise2007}
G.~P{\'e}piot, N.~Cheikhrouhou, J.-M. Furbringer, and R.~Glardon, ``{{UECML}}:
  {{Unified Enterprise Competence Modelling Language}},'' \emph{Computers in
  Industry}, vol.~58, pp. 130--142, 2007.

\bibitem{elasameCompetencyModelReview2018}
M.~El~Asame and M.~Wakrim, ``\BIBforeignlanguage{en}{Towards a competency
  model: {{A}} review of the literature and the competency standards},''
  \emph{\BIBforeignlanguage{en}{Education and Information Technologies}},
  vol.~23, no.~1, pp. 225--236, Jan. 2018.

\bibitem{conatiBayesianStudentModeling2010}
C.~Conati, ``\BIBforeignlanguage{en}{Bayesian {{Student Modeling}}},'' in
  \emph{\BIBforeignlanguage{en}{Advances in {{Intelligent Tutoring Systems}}}},
  ser. Studies in {{Computational Intelligence}}.\hskip 1em plus 0.5em minus
  0.4em\relax {Springer}, 2010, no. 308, pp. 281--299.

\bibitem{kaserDynamicBayesianNetworks2017}
T.~K{\"a}ser, S.~Klingler, A.~G. Schwing, and M.~Gross,
  ``\BIBforeignlanguage{en}{Dynamic {{Bayesian Networks}} for {{Student
  Modeling}}},'' \emph{\BIBforeignlanguage{en}{IEEE Transactions on Learning
  Technologies}}, vol.~10, no.~4, pp. 450--462, 2017.

\bibitem{sucarProbabilisticGraphicalModels2015a}
L.~E. Sucar, \emph{\BIBforeignlanguage{en}{Probabilistic {{Graphical Models}}
  \textendash{} {{Principles}} and {{Applications}}}}, ser. Advances in
  {{Computer Vision}} and {{Pattern Recognition}}.\hskip 1em plus 0.5em minus
  0.4em\relax {London}: {Springer}, 2015, no. 4205.

\bibitem{chanCompetencyAnalyserKnowledgebased2010}
M.~E. Chan, S.~C. Gonz{\'a}lez, and R.~Morales, ``A competency analyser as a
  knowledge-based approach for making e-learning more flexible and
  personalised,'' in \emph{{{EDULEARN10 Proceedings CD}}. 2nd {{International
  Conference}} on {{Education}} and {{New Learning Technologies}}
  ({{Edulearn}})}, {Barcelona}, 2010, pp. 1607--1612.

\bibitem{medinafloresMapaCompetenciasGenericas2017}
R.~Medina~Flores and R.~Morales~Gamboa, ``\BIBforeignlanguage{es}{{Mapa de
  competencias gen{\'e}ricas del Acuerdo 444 de la RIEMS para entornos
  virtuales inteligentes basados en competencias}},'' in
  \emph{\BIBforeignlanguage{es}{{Octavo Coloquio Nacional de Educaci{\'o}n
  Media Superior a Distancia}}}.\hskip 1em plus 0.5em minus 0.4em\relax
  {Le{\'o}n, Guanajuato}: {Universidad Virtual del Estado de Guanajuato}, 2017.

\bibitem{secretariadeeducacionpublicaACUERDONumero4442008}
{Secretar{\'i}a de Educaci{\'o}n P{\'u}blica}, ``{{ACUERDO}} n{\'u}mero 444 por
  el que se establecen las competencias que constituyen el marco curricular
  com{\'u}n del {{Sistema Nacional}} de {{Bachillerato}},'' \emph{Diario
  Oficial de la Federaci{\'o}n}, Oct. 2008.

\bibitem{holtStateStudentModelling1994}
P.~Holt, S.~Dubs, M.~Jones, and J.~Greer, ``The {{State}} of {{Student
  Modelling}},'' in \emph{Student {{Modelling}}: {{The Key}} to
  {{Individualized Knowledge}}-{{Based Instruction}}}, J.~Greer and G.~I.
  McCalla, Eds.\hskip 1em plus 0.5em minus 0.4em\relax {Springer Verlag}, 1994,
  vol. 125, pp. 3--35.

\bibitem{vanlehnStudentModelling1988}
K.~VanLehn, ``Student {{Modelling}},'' in \emph{Foundations of {{Intelligent
  Tutoring Systems}}}, M.~C. Polson and J.~J. Richardson, Eds.\hskip 1em plus
  0.5em minus 0.4em\relax {New Jersey}: {Lawrence Erlbaum Associates}, 1988,
  pp. 55--78.

\bibitem{murphyDynamicBayesianNetworks2002}
K.~P. Murphy, ``\BIBforeignlanguage{en}{Dynamic {{Bayesian Networks}}:
  {{Representation}}, {{Inference}} and {{Learning}}},'' Ph.{{D}}. {{Thesis}},
  University of California at Berkeley, 2002.

\bibitem{vygotskyMindSocietyDevelopment1978}
L.~S. Vygotsky and M.~Cole, \emph{Mind in {{Society}}: {{The Development}} of
  {{Higher Psychological Processes}}}.\hskip 1em plus 0.5em minus 0.4em\relax
  {Harvard University Press}, 1978.

\end{thebibliography}

\end{document}